\documentclass[11pt]{article}
\usepackage{latexsym}
\usepackage{graphicx}
\usepackage{caption2}
\usepackage{psfrag}
\usepackage{amsmath}
\usepackage{setspace}
\newcommand{\be}{\begin{equation}}
\newcommand{\ee}{\end{equation}}
\newcommand{\bea}{\begin{eqnarray}}
\newcommand{\eea}{\end{eqnarray}}

\newcommand{\lesssim}{ {\
\lower-1.2pt\vbox{\hbox{\rlap{$<$}\lower5pt\vbox{\hbox{$\sim$}}}}\ } }
\newcommand{\gtrsim}{ {\
\lower-1.2pt\vbox{\hbox{\rlap{$>$}\lower5pt\vbox{\hbox{$\sim$}}}}\ } }

\input epsf

\begin{document}

\begin{titlepage}

\begin{flushright}
\end{flushright}
\vspace*{1.5cm}
\begin{center}
{\Large \bf A holographic approach to low-energy weak interactions of hadrons}
\\[2.0cm]

{\bf L. Cappiello}$^{a,b}$, {\bf O. Cat\`{a}}$^{c}$ and {\bf G. D'Ambrosio}$^{b}$\\[1cm]

 $^{a}$Dipartimento di Scienze Fisiche,
             Universit\'a di Napoli "Federico II", Via Cintia, 80126 Napoli, Italy\\[.5 cm]
 $^{b}$INFN-Sezione di Napoli, Via Cintia, 80126 Napoli, Italy\\[.5 cm]
 $^{c}$Departament de F\'isica Te\`orica and IFIC, Universitat de Val\`encia-CSIC,
Apartat de Correus 22085, E-46071 Val\`encia, Spain

\end{center}

\vspace*{1.0cm}

\begin{abstract}
We apply the double-trace formalism to incorporate nonleptonic weak interactions of hadrons into holographic models of the strong interactions. We focus our attention upon $\Delta S=1$ nonleptonic kaon decays. By working with a Yang-Mills--Chern-Simons 5-dimensional action, we explicitly show how, at low energies, one recovers the $\Delta S=1$ weak chiral Lagrangian for both the anomalous and nonanomalous sectors. We provide definite predictions for the low energy coefficients in terms of the AdS metric and argue that the double-trace formalism is a 5-dimensional avatar of the Weak Deformation Model introduced long ago by Ecker et al. As a significant phenomenological application, we reassess the $K\to 3\pi$ decays in the light of the holographic model. Previous models found a fine-tuned cancellation of resonance exchange in these decays, which was both conceptually puzzling and quantitatively in disagreement with experimental results. The holographic model we build is an illustrative counterexample showing that the cancellation encountered in the literature is not generic but a model-dependent statement and that agreement with experiment can be obtained.
\end{abstract}

\end{titlepage}

\section{Introduction}\label{secI}

The AdS/CFT correspondence conjectured by Maldacena~\cite{Maldacena:1997re} is one of the holographic dualities with more far-reaching implications for gauge theories. Subsequent developments~\cite{Gubser:1998bc,Witten:1998qj} provided the necessary tools to make this holographic duality quantitative, to the point that nowadays it stands as one of the most solid approaches to the study of strongly-coupled theories. The first application to the strong interactions was made in the deconstructed model of~\cite{Son:2003et}, inspired in Hidden Local Symmetry models~\cite{Bando:1984ej}. Soon thereafter versions with a continuum fifth dimension appeared~\cite{Erlich:2005qh,Sakai:2004cn}. There are many features that make holographic approaches to QCD attractive. In the first place, the AdS metric endows the model with conformal symmetry. As a result, models automatically exhibit the short distances of the parton model. Moreover, the low energy limit of the theory, {\it{i.e.}}, chiral perturbation theory, is easily reached once chiral symmetry is implemented. With these two ingredients, the theory is guaranteed to smoothly interpolate between long and short distance QCD. Additionally, since dimensional reduction generates an infinite number of resonances (the Kaluza-Klein excitations), holographic theories can be viewed as realizations of QCD in the large-$N_c$ limit. This viewpoint was adopted in~\cite{Hirn:2005nr}, where the phenomenology of vector and axial-vector QCD correlators was investigated.

However, strong effects not only affect QCD correlators, but are also present in correlators involving weak currents. In order to obtain predictions for nonperturbative electroweak parameters, the previous framework should be extended to include weak interactions of hadrons. A first step in this direction was taken in~\cite{Hambye:2005up}. Here we will follow a different approach. In chiral perturbation theory~\cite{Gasser:1983yg}, the electroweak interactions are introduced as a perturbation to the strong Lagrangian through the method of external sources. At the holographic level there exists a prescription to include perturbative effects in the form of multiple-trace operators~\cite{Witten:2001ua}. Since at low energies the weak interactions adopt a current-current structure, it was recently suggested~\cite{Gazit:2008gz} that electroweak effects could be incorporated in holographic models as double-trace deformations.

In this paper we will apply the previous ideas to nonleptonic kaon decays. These are $\Delta S=1$ processes mediated by W exchange which, at energies below the W boson mass, can be described by the effective Hamiltonian~\cite{Gilman:1979bc}
\begin{equation}\label{GW}
{\cal H}_{eff}^{|\Delta S|=1}=-\kappa Q_{-}+h.c.~,\qquad \kappa=-\frac{G_F}{\sqrt{2}}V_{ud} V_{us}^*C_{-}(\mu)~,
\end{equation}
where 
\begin{equation}
Q_{-}=4(\overline{s}_L\gamma^{\mu}u_L)(\overline{u}_L\gamma_{\mu}d_L)-4(\overline{s}_L\gamma^{\mu}d_L)(\overline{u}_L\gamma_{\mu}u_L)~,
\label{eq:qplusm}
\end{equation}
is a four-quark current-current operator of the light $u,d,s$ quarks and $C_-(\mu)$ is a Wilson coefficient that collects both the integrated degrees of freedom above the energy-scale $\mu$ as well as perturbative QCD corrections. At energies of the order of the kaon mass, quarks and gluons hadronize and the proper framework is chiral perturbation theory, where one requires the bosonized version of Eq.~(\ref{GW}). 

In order to describe nonleptonic kaon decays from an holographic perspective we will start from a strong 5-dimensional Yang-Mills--Chern-Simons action and introduce Eq.~(\ref{GW}) with the bosonized $Q_-$ as a double-trace perturbation. We will explicitly show that at low energies one recovers the well-known chiral Lagrangian for the strong and $\Delta S=1$ electroweak interactions, with definite predictions for the low-energy couplings. Interestingly, the double-trace formalism turns out to be the 5-dimensional analog of the so-called Weak Deformation Model (WDM), a heuristic model introduced in~\cite{EckerWDM} to constrain the weak low-energy couplings. We clarify the relation between WDM and factorization in 4-dimensional theories by showing that the former is a truncated version of the latter, and discuss the difficulties to achieve full factorization in holographic settings.

As a representative phenomenological application, we will discuss $K\to 3\pi$ decays. These modes were examined in the past using different hadronic models~\cite{Isidori:1991ya,Ecker:1992de,D'Ambrosio:1997tb}. One of the underlying assumptions in all these models (and also in our holographic prescription) is that (vector) resonance exchanges are the bulk contribution to {\emph{both}} strong and electroweak chiral couplings. This {\emph{resonance saturation}} was shown to be successful for the strong sector~\cite{Ecker:1988te} and for many weak channels, but failed dramatically for $K\to 3\pi$ decays. Since all the existing models predicted a vanishing resonance contribution, $K\to 3\pi$ has stood out as an unexplained exception to resonance saturation. We show that the vanishing resonance contribution is not a generic feature of $K\to 3\pi$ but it is due to a model-dependent accidental cancellation, which is phenomenologically disfavored. Interestingly, the holographic model does not reproduce such cancellation, better phenomenological agreement is achieved and shows that resonance saturation is at work also for this channel. 

This paper is organized as follows: In Section~\ref{secII} we will introduce the holographic model for the strong interactions. The multiple-trace formalism will be briefly discussed in Section~\ref{secIII} and applied to nonleptonic kaon decays. In Section~\ref{secIV} we work out the low-energy limit of the model and recover the chiral Lagrangian for the strong and electroweak theories to next-to-leading order, together with predictions for the low-energy couplings. The anomalous sector, related to the Chern-Simons term, is analyzed in Section~\ref{secV}. Section~\ref{secVI} is devoted to $K\to 3\pi$ decays within the holographic model. Conclusions are given in Section~\ref{secVII}.    


\section{The holographic model}\label{secII}
We will work with the model of Ref.~\cite{Hirn:2005nr}, given by the chiral invariant 5-dimensional Yang-Mills action
\begin{equation}
S_{\rm YM}[L_M,R_M]=-\frac{1}{4g_5^2}\, \int d^4x \int_0^{z_0} dz~
\sqrt{g}
\left\langle F_{(L)}^{MN}F_{(L)MN}+F_{(R)}^{MN}F_{(R)MN}\right\rangle~.\label{definitions}
\end{equation} 
In order to account for the anomalous sector of the theory, described by parity-odd operators, we will also consider the 5-dimensional Chern-Simons action:  
\begin{equation}
S_{{\mathrm{CS}}}[L_M,R_M]=\frac{N_c}{24\pi^2}\int_{\mathrm{AdS_5}} \left[\omega_5(L_M)-\omega_5(R_M)\right]~,
\end{equation}
where
\begin{equation}
\omega_5(L_M)=\left\langle L F_{(L)}^2-\frac{1}{2}L^3 F_{(L)}+\frac{1}{10}L^5\right\rangle~. 
\end{equation}
In the previous equations $\langle\cdots\rangle$ stands for the trace over flavor, $L_M=L_M^aT^a$ is the $SU(3)_{L}$ gauge field, {\it{i.e.}}, $L_M\to G_LL_MG_L^{\dagger}+iG_L\partial_M G_L^{\dagger}$ and, accordingly, $F_{(L)MN} =\partial_{M}{L}_{N} -\partial_{N} {L}_{M}-i[L_{M},L_{N}]$. For the Chern-Simons form we use the conventions $F_{(L)}=dL+L^2$ with $L=-iL^aT^a$. Similar expressions hold for the right-handed field. Their relation to vector and axial fields is $V_M=R_M+L_M$ and $A_M=R_M-L_M$. We choose the metric to be pure AdS
\begin{align}
g_{MN}dx^Mdx^N = z^{-2}\left(\eta_{\mu \nu}dx^{\mu}dx^{\nu} -
dz^2\right)~,\label{metricwarp}
\end{align}
where $\eta_{\mu\nu}={\rm Diag}\,(1,-1,-1,-1)$, $\mu,\nu=(0,1,2,3)$ and $M,N=(0,1,2,3,z)$, over a finite interval $(0,z_0]$. As a result, boundary conditions for the fields have to be specified. On the UV brane the AdS/CFT correspondence prescribes
\begin{equation}
L_{\mu}(x,0)=l_\mu(x)~, \quad
R_{\mu}(x,0)=r_\mu(x)~,\label{UV}
\end{equation}
where $l_\mu(x)$ and $r_\mu(x)$ are identified with the classical 4-dimensional sources coupled to the chiral currents $J_{L\,\mu}= \overline{q}_L\gamma^{\mu}q_L$ and $J_{R\,\mu}= \overline{q}_R\gamma^{\mu}q_R$. They transform as $l_{\mu}\to g_L l_{\mu}g^{\dagger}_L+ig_L\partial_{\mu}g^{\dagger}_L$ (similarly for $r_{\mu}$ with the obvious replacements), where $g_L$ is the restriction of $G_L$ on the UV brane. 

While conditions on the UV brane come naturally from the holographic duality between gravity and gauge theories, the choice of boundary conditions on the IR brane is dictated by the low energy characteristics of the field theory under study. In the case of QCD, the presence of the IR brane itself guarantees that conformal invariance is broken and generates a hadronic spectrum with an infinite number of resonances. Chiral symmetry and its breaking pattern can be implemented in different ways. In Ref.~\cite{Hirn:2005nr} chiral symmetry breaking was induced entirely through the IR boundary conditions. In order to reproduce the observed pattern $SU(3)_L\times SU(3)_R\to SU(3)_V$ it is convenient to work with Dirichlet IR boundary conditions for the axial field and Neumann boundary conditions for the vector field. In terms of the left and right chiral fields they read    
\begin{align}
L_{\mu}(x, z_0) -R_{\mu}(x,z_0)=0~,&\label{IRbreak}\\
F_{(L)}^{z\mu}(x,z_0) +F_{(R)}^{z\mu}(x,z_0)=0~.&\label{IRnobreak}
\end{align}
The previous asymmetric choice of boundary conditions ensures that: (a) chiral symmetry is broken and thereby an energy splitting between vector and axial resonances is generated; (b) the Neumann boundary condition for the vector field is gauge invariant and therefore $SU(3)_V$ is a symmetry of the field theory;  while (c) Dirichlet boundary conditions break gauge invariance and lead to the appearance of an axial zero-mode.

The axial zero-mode can then be interpreted as the pion in the following way. Since the 5-dimensional fields are massless, there is some gauge redundancy that can be eliminated. For convenience one works in the axial gauge, such that $L_{z}=R_z=0$. This can be achieved by gauge-transforming the 5-dimensional fields with the following Wilson lines
\begin{equation}
\xi_{L}(x,z)=P  \exp \left\{-i \int_z^{z_0} dz'\,
L_z(x,z') \right\}~,\quad \xi_{R}(x,z)=P  \exp \left\{-i \int_z^{z_0} dz'\,
R_z(x,z') \right\}~,\label{Wilson}
\end{equation}
defined such that they start at the IR brane and end at the UV brane. Thus, $\xi_{L,R}(x,z_0)=0$ by construction and the infrared boundary conditions Eq.~(\ref{IRbreak}) are respected.\footnote{Notice that Eq.~(\ref{IRnobreak}) is trivially satisfied due to gauge invariance.} In contrast, $\xi_{L,R}(x,0)\neq 0$ and the UV boundary conditions of Eq.~(\ref{UV}) change to the chirally-dressed expressions ($\xi_{\Lambda}(x,0)\equiv \xi_{\Lambda}(x)$):  
\begin{align}
L_{\mu}^{(0)}(x)=\xi_L^\dagger(x)[l_\mu(x)+i\partial_\mu]\xi_L(x)~,\label{UVchiralL}
\\
R_{\mu}^{(0)}(x)=\xi_R^\dagger(x)[r_\mu(x)+i\partial_\mu]\xi_R(x)~.\label{UVchiralR}
\end{align}
The Wilson lines satisfy $\xi_{\Lambda}(x)\to g_{\Lambda}(x) \xi_{\Lambda}(x) h(x)$, where $g_{\Lambda}\in SU(3)_{\Lambda}$ and $h(x)\in SU(3)_V$. One can then eliminate the residual dependence on $h(x)$ by building the $SU(3)_L\times SU(3)_R$ invariant object:
\begin{align}
U(x)=\xi_{R}(x)\xi_{L}^\dagger (x)~,\label{Uchiral}
\end{align}
which is a chiral field that contains the Goldstone bosons. It is common to work in the particular gauge $\xi_R(x)=\xi_L^{\dagger}(x)\equiv u(x)$. From here on we will adopt this gauge. Therefore,
\begin{align}
L_{\mu}^{(0)}(x)&=u\left(l_\mu + i \partial_\mu\right)u^\dagger =i\Gamma_{\mu}-\frac{1}{2}u_{\mu}~,\nonumber\\
R_{\mu}^{(0)}(x)&=u^\dagger\left(r_\mu + i \partial_\mu\right)u =i\Gamma_{\mu}+\frac{1}{2}u_{\mu}~,\label{chiraldressing}
\end{align}
where $U=u^2$ and we have used the chiral connection $\Gamma_{\mu}$ and the vielbein $u_{\mu}$, defined as
\begin{align}
\Gamma_{\mu}&=\frac{1}{2}\left[u^{\dagger}(\partial_{\mu}-ir_{\mu})u+u(\partial_{\mu}-il_{\mu})u^{\dagger}\right]~,\\
u_{\mu}&=i\left[u^{\dagger}(\partial_{\mu}-ir_{\mu})u-u(\partial_{\mu}-il_{\mu})u^{\dagger}\right]=iu^{\dagger}D_{\mu}U u^{\dagger}=-iuD_{\mu}U^{\dagger}u~,
\end{align}
with $D_{\mu}U=\partial_{\mu}U-ir_{\mu}U+iUl_{\mu}$. $\Gamma_{\mu}$ and $u_{\mu}$ are chirally-dressed vector and axial source fields, respectively, {\it{i.e.}}, $V_{\mu}^{(0)}(x)=2i\Gamma_{\mu}$ and $A_{\mu}^{(0)}(x)=u_{\mu}$. In view of what we will discuss in subsequent sections it is convenient to examine the behavior of the on-shell solutions of Eq.~(\ref{definitions}) away from the UV boundary. Solving the equations of motion one finds
\begin{align} 
L_\mu (x,z)&= L^{(0)}_\mu (x)+L^{(1)}_\mu(x)z^2+{\hat{L}}(x,z)~,\nonumber\\ 
R_\mu (x,z)&= R^{(0)}_\mu (x)+R^{(1)}_\mu(x)z^2+{\hat{R}}(x,z)~.\label{smallz}
\end{align} 
The first two term in each equation are the solution of the linearized equations of motion at zero momentum, {\it{i.e.}}, they describe the zero modes of the bulk-to-boundary propagators. The quadratic dependence on $z$ is a direct consequence of the conformal invariance induced by the AdS metric. Taking into account the boundary conditions of Eqs.~(\ref{IRbreak}) and (\ref{IRnobreak}), one can show that 
\begin{align}
L^{(1)}_\mu =- R^{(1)}_\mu =\frac{1}{2z_0^2}u_{\mu}~.\label{JL1R1}
\end{align}
The last terms in Eqs.~(\ref{smallz}) encode the contributions of the Kaluza-Klein tower of massive resonances, which can be found by solving the equations of motion for non-zero momentum. We will not give their explicit expressions. For our purposes it will suffice to note that their contribution starts at ${\cal{O}}(p^3,z^3)$. 


\section{Effective weak Hamiltonian as a double-trace deformation of holographic QCD}\label{secIII}

In this section we will briefly summarize the discussion of Ref.~\cite{Witten:2001ua} for multiple-trace operators to later on apply it to the electroweak theory.

\subsection{Multiple-trace operators in AdS/CFT}
Let us consider a generic field $\phi(x,z)$ in AdS space. Near the UV boundary, $z\to 0$, the solution of the free equation of motion is
\begin{equation} 
\phi(x,z)\sim \phi_{0}(x)  z^{\Delta_-}+\phi_{1}(x)
z^{\Delta_+}~,\label{nearUV}
\end{equation}
where $\Delta_{\pm}$ are the roots of the equation $\Delta(\Delta+d)=m_{\phi}^2$, where $m_{\phi}$ is the 5-dimensional mass of the scalar field. If we choose $\Delta_+>\Delta_-$, then $\phi_0(x)$ is the leading coefficient while $\phi_1(x)$ is subleading. The evaluation of the 5-dimensional action with the field $\phi(x,z)$ on-shell leaves a UV boundary term, which according to the AdS/CFT prescription for correlators~\cite{Gubser:1998bc,Witten:1998qj}
\begin{equation}
\mbox{exp}\left(i
S_5[\phi_0(x)]\right)=\langle\mbox{exp}\left[i\int d^4 x
\,s(x)\,{\cal O}(x)\right]\rangle_{\rm
QCD_4}~,\label{holography}
\end{equation}
is identified with the generating functional of the 4-dimensional theory in the presence of an external (classical) source $s(x)$ coupled to the single-trace operator ${\cal O}(x)$ (with conformal dimension $\Delta_-$). Quite generically, one finds that
\begin{equation}
W=\int d^4 x\,s(x)\,{\cal O}(x)\sim\int d^4x \phi_0(x)\phi_1(x)~.\label{generic}
\end{equation}
The previous prescription allows one to identify $\phi_0(x)$ and $\phi_1(x)$ as, respectively, the source and the one-point function for the operator ${\cal{O}}(x)$. More formally, one can write\footnote{Depending on the field under consideration, the expressions above may need to be regularized. In that case, one should replace $W\to W+W_c$, where $W_c$ contains operators acting as counterterms. In this paper we will be dealing with vector fields, for which no such regularization is needed.} 
\begin{align}
&\phi_{0}(x)=s(x)=\frac{\delta W}{\delta \phi_1}~,\label{0-0}\\
&\phi_{1}(x)=\langle {\cal O}(x)\rangle_{s}=\frac{\delta W}{\delta \phi_0}~,\label{0-1}
\end{align}
which explicitly shows that $\phi_0(x)$ and $\phi_1(x)$ are canonically conjugated quantities.
     
Strictly speaking the identification~\eqref{0-1} can be fully exploited only after the IR behavior of the solution has been fixed, either by the requirement of normalizability, in the pure AdS case, or by suitable IR boundary conditions in QCD-like models. Once this is done, $\phi_1(x)$ can be related to $\phi_0(x)$ through 
\begin{equation}
\phi_1(x)=\int\,d^4\,x' G(x,x')\,\phi_0(x)~,
\end{equation}
and accordingly
\begin{equation}
W=\int\,d^4\,x\,d^4\,x's(x)\, G(x,x')\,s(x')~,
\end{equation}
where $G(x,x')$ is the two-point function in coordinate space. In other words, once IR boundary conditions are imposed the $z$-dependent factor builds the bulk-to-boundary propagator. 
 
So far the discussion has been restricted to single-trace operators ${\cal{O}}(x)$. Perturbations due to multi-trace operators will generate a functional $W[{\cal{O}}]$ no longer linear in ${\cal{O}}$. The prescription outlined in~\cite{Witten:2001ua} is to identify the sources for this more general case by imposing that the canonical relations of Eqs.~(\ref{0-0},\ref{0-1}) remain valid for arbitrary $W[{\cal{O}}]$. Therefore, 
\begin{equation}
\phi_0(x)=\frac{\delta W[{\cal O}]}{\delta
{\cal O}}\Bigg|_{{\cal O}\to
\phi_1(x)}~.\label{prescription}
\end{equation}
For our purposes we will be interested in double-trace perturbations of the form $W[{\cal{O}}]=W_0[{\cal{O}}]+\zeta_1W_1[{\cal{O}}^2]$ and therefore the previous prescription will imply the (infinitesimal) canonical transformation generated by $W_1[{\cal{O}}]$:
\begin{equation}\label{prescW}
\langle{\cal{O}}(x)\rangle_s=\phi_1(x)~,\qquad s(x)=\phi_0(x)+\zeta_1\frac{\delta W_1[{\cal{O}}]}{\delta {\cal O}}\Bigg|_{{\cal O}\to
\phi^{(1)}(x)}~.
\end{equation} 


\subsection{Double-trace formalism for the weak interactions}

Let us now apply the general procedure outlined previously to introduce weak sources in the holographic model of Section~\ref{secII}. Let us first determine the chiral currents. For that we only need to consider $W_0[J_{L\mu},J_{R\mu}]$, which can be easily found to be
\begin{equation}
W_0[J_{L\mu},J_{R\mu}]=-\frac{1}{2g_5^2}\lim_{z\to 0}\int d^4x\langle L_{\mu}\frac{1}{z}\partial_z L^{\mu}+R_{\mu}\frac{1}{z}\partial_z R^{\mu}\rangle~.\label{sources} 
\end{equation}
Plugging Eq.~(\ref{smallz}) into the previous expression, one finds that $W_0[J_{L\mu},J_{R\mu}]\sim\int d^4x~\eta^{\mu\nu}\langle L^{(0)}_{\mu}(x)L^{(1)}_{\nu}(x)+R^{(0)}_{\mu}(x)R^{(1)}_{\nu}(x)\rangle$, which complies with Eq.~(\ref{generic}). Matching it to the general form $W_0[J_{L\mu},J_{R\mu}]=\int d^4x\langle l^{\mu}(x)J_{L \mu}(x)+r^{\mu}(x)J_{R \mu}(x)\rangle$, one can extract the chiral currents as the derivatives over the source fields. Proceeding this way one finds
\begin{align}
\langle
J_{L\,\mu}(x)\rangle_{l_\mu}&=-\frac{f_\pi^2}{2}u^\dagger
u_{\mu}u=-i \frac{f_\pi^2}{2}U^\dagger D_{\mu} U~,
\label{JL2-1}
\\
\langle J_{R\,\mu}(x)\rangle_{l_\mu}&=\frac{f_\pi^2}{2}u
u_{\mu}u^\dagger=i\frac{f_\pi^2}{2}UD_{\mu}
U^\dagger~.\label{JL2-2}
\end{align}
As expected, the previous expressions correspond to the bosonized chiral currents of the ${\cal O}(p^2)$ strong chiral Lagrangian, where we have identified $f_{\pi}=\sqrt{2}(g_5z_0)^{-1}$. Incidentally, notice that the chiral dressing of the fields $L^{\mu}(x)$ and $R^{\mu}(x)$ carries over to the current operators. We stress that the previous expressions refer to one-point functions of 4-dimensional operators with external (chiral) sources turned on. Obviously, they are non-vanishing even when the external sources are turned off. In that case, they are vacuum expectation values of S$\chi$SB induced by the IR boundary conditions.

Let us now consider the addition of electroweak $\Delta S=1$ nonleptonic operators. As we discussed in the Introduction, at energy-scales $\Lambda_{QCD}\lesssim \mu \lesssim m_W$ the effective Hamiltonian is proportional to a single operator, $Q_-$. Therefore, the effective action on the UV boundary will take the form $W[J_{L\mu},J_{R\mu}]=W_0[J_{L\mu},J_{R\mu}]+\kappa W_1[J_{L\mu}]$, with $\kappa$ defined in Eq.~(\ref{GW}) and $W_1[J_{L\mu}]$ proportional to the chiral realization of $Q_-$. The resulting operators transform under the gauge group as $(8_L,1_R)$ and $(27_L,1_R)$, which correspond, respectively, to $\Delta I=1/2$ and $\Delta I=3/2$ transitions. It turns out that phenomenologically the octet operators are enhanced, a circumstance known as the $\Delta I=1/2$ rule. Thus, to a very good approximation, the bosonization of $Q_-$ amounts to the replacement\footnote{We are neglecting CP-violating terms. If wanted, they can always be reinstated by replacing $\lambda_6\to \frac{1}{2}(\lambda_6-i\lambda_7)$ and adding the hermitian conjugate.}      
\begin{equation}
Q_-\to \langle\lambda_6 J_{L\mu}J_L^{\mu}\rangle-\langle\lambda_6 J_{L\mu}\rangle\langle J_L^{\mu}\rangle~,
\end{equation}
and as a result $W_1[J_{L\mu}]$ is given by
\begin{align}
W_1[J_{L\mu}]=\zeta_8\int d^4x \left\{\langle\lambda_6 J_{L\mu}J_L^{\mu}\rangle-\langle\lambda_6 J_{L\mu}\rangle\langle J_L^{\mu}\rangle\right\}~,
\end{align}
where $\zeta_8$ is a chiral coupling that accounts for nonperturbative effects below $\Lambda_{QCD}$.

Direct application of Eqs.~(\ref{prescW}) results in the following shift in the left-handed sources:
\begin{align} 
l_\mu\rightarrow l_{\mu}+\kappa \frac{\delta W_1[J_{L\mu}]}{\delta J_{L\,
\mu}}&=l_\mu +\kappa\zeta_8\bigg[\{\lambda_6,J_{L\,
\mu}\}-\textbf{1}_3\langle\lambda_6J_{L\,
\mu}\rangle-\langle J_{L\, \mu}\rangle\,
\lambda_6\bigg]\Bigg|_{Eq.~(\ref{JL2-1})}\nonumber\\
&=l_\mu -\kappa\zeta_8\frac{f^2_\pi}{2}\,\left(\{\lambda_6,u^\dagger
u_\mu u\}-\textbf{1}_3\langle\lambda_6\, u^\dagger u_\mu
u\rangle\right)~. \label{lmudeformed}\end{align} 
In the last line we have dropped the term $\langle u^\dagger u_\mu u\rangle$ since the chiral field is traceless. The term proportional to the identity is a $1/N_C$-suppressed operator that comes from the nonet component of the weak interaction.

In terms of the chirally-dressed fields at the UV boundary, the shift $l_{\mu}\to l_{\mu}+\kappa \delta l_{\mu}^W$ in Eq.~(\ref{lmudeformed}) amounts to $L_{\mu}^{(0)}\to L_{\mu}^{(0)}+\kappa \ell_{\mu}^W$, with
\begin{equation}
\ell_{\mu}^{W}=-\zeta_8\frac{f_\pi^2}{2}\left(\{\Delta, u_\mu\}-
 \textbf{1}_3\langle\Delta\,u_\mu
\rangle\right)~,\qquad \Delta\equiv u\,\lambda_6 \,u^\dagger~,\label{definition}
\end{equation}
written in such a way that every object transforms in the adjoint representation of $SU(3)_V$. 
 
Notice that the previous prescription allows one to obtain the weak Lagrangian to all orders in the chiral expansion (provided the action is consistently improved with higher-order operators) through the recipe $S[U,l,r]\to S[U,l,r]+\kappa \delta S[U,l,r]$, where 
\begin{equation}
\delta S[U,l,r]=\frac{\delta S[U,l,r]}{\delta l_{\mu}}\delta l_{\mu}^W~.
\end{equation}
Therefore, up to ${\cal{O}}(p^4)$ in the chiral expansion one needs
\begin{equation}
\delta S[U,l,r]=\left[\frac{\delta S_2}{\delta l_{\mu}}+\frac{\delta S_4}{\delta l_{\mu}}+\frac{\delta S_{WZW}}{\delta l_{\mu}}\right]\delta l_{\mu}^W= \left[J_{L}^{\mu\,(1)}+J_{L}^{\mu\,(3)}+J_{L}^{\mu\,WZW}\right]\delta l_{\mu}^W~,\label{curr}
\end{equation}
where $\frac{\delta{S_i}}{\delta l_{\mu}}\equiv J_{L}^{\mu\,(i-1)}$ are the associated chiral currents.


\section{Low energy regime}\label{secIV}
The starting point is the 5-dimensional action in the axial gauge $L_z=R_z=0$, which can be expressed as
\begin{align}
 S = -\frac{1}{4 g_5^2} \int d^4 x \int_{0}^{z_0}\frac{dz}{z} &\left\langle-2\,\eta^{\mu\,\nu}\,\left( \partial_z L_\mu
\partial_z L_\nu+
\partial_z R_\mu \partial_z R^\mu\right) \nonumber\right.\nonumber\\
&\left. +\eta^{\mu\,\rho}\,\eta^{\nu\,\sigma}
\left(F_{(L)\,\mu\nu} F_{(L)\,\rho\sigma}+ F_{(R)\,\mu\nu}
F_{(R)\,\rho\sigma}\right)\right\rangle~. \label{5Dgfaction}
\end{align}
In order to extract the different contributions in the chiral expansion it is convenient to decompose the fields in the zero-mode and resonance contribution, as we did in Eq.~(\ref{smallz}). From the expressions found for $L_{\mu}^{(0)}$ and $R_{\mu}^{(0)}$ is it obvious that they are ${\cal{O}}(p)$, while the resonance pieces are ${\cal{O}}(p^3)$. Thus, in order to extract the leading ${\cal{O}}(p^2)$ chiral contribution from Eq.~(\ref{5Dgfaction}) one needs to consider only the zero-mode contributions in the first line of Eq.~(\ref{5Dgfaction}). This is in full agreement with the expectation that the leading pieces in the chiral Lagrangian are universal, {\it{i.e.}}, independent of the resonance model and only based on the pattern of chiral symmetry breaking. Two comments are in order at this point: (i) our results will be restricted to the chiral limit; and (ii) we will limit ourselves to ${\cal{O}}(p^4)$ contributions. Beyond that order, consistency would require to include higher-order operators into the action.


\subsection{Leading ${\cal{O}}(p^2)$ operators}
The leading strong and electroweak chiral Lagrangians are commonly parametrized as
\begin{align}\label{chiral}
S_2=\int d^4x~\frac{f_{\pi}^2}{4}\langle u_{\mu}u^{\mu}\rangle;\qquad S_2^W=\int d^4x~G_8 f_{\pi}^4\langle \Delta u_{\mu}u^{\mu} \rangle~.
\end{align}
From the holographic point of view the quantity we have to evaluate can be expressed as
\begin{align}\label{holop2}
S_2&= -\frac{1}{4 g_5^2} \int d^4 x \int_{0}^{z_0} \frac{dz}{z}\left\langle -2\,\eta^{\mu\,\nu}\,\left( \partial_z L_\mu
\partial_z L_\nu+
\partial_z R_\mu \partial_z R^\mu\right) \nonumber\right\rangle\nonumber\\
&=-\frac{1}{2 g_5^2} \int d^4 x \langle L_\mu \frac{1}{z}\partial_z L^\mu+R_\mu \frac{1}{z}\partial_z R^\mu\rangle\Bigg|_{z\to 0}\nonumber\\
&=-\frac{1}{g_5^2}\int d^4 x \langle L_{\mu}^{(0)}L^{\mu{(1)}}+R_{\mu}^{(0)}R^{\mu{(1)}}\rangle~,
\end{align}
where we have integrated by parts and used Eq.~(\ref{smallz}). The strong Lagrangian can be readily found using the results of Eqs.~(\ref{chiraldressing}) and (\ref{JL1R1}). The result only involves the axial component of the fields, as it should, and one recovers the familiar expression:
\begin{align} 
S_2&=\frac{1}{2 g_5^2z_0^2}\int d^4 x \langle u_{\mu}u^{\mu}\rangle~.
\end{align}
Matching to the strong part of Eq.~(\ref{chiral}) requires that $f_{\pi}=\sqrt{2}(g_5z_0)^{-1}$, which is consistent with the identification we made when discussing the chiral currents. The leading electroweak term can be found by shifting the fields in Eq.~(\ref{holop2}) as 
\begin{align}
L_{\mu}^{(0)}&\to L_{\mu}^{(0)}+\kappa\ell_{\mu}^W~;&R_{\mu}^{(0)}\to &R_{\mu}^{(0)}~;\nonumber\\
L_{\mu}^{(1)}&\to L_{\mu}^{(1)}-\frac{\kappa}{2z_0^2}\ell_{\mu}^W~; &R_{\mu}^{(1)}\to &R_{\mu}^{(1)}+\frac{\kappa}{2z_0^2}\ell_{\mu}^W~.
\end{align}
Notice that, according to Eq.~(\ref{JL1R1}), $\emph{both}$ $L_{\mu}^{(1)}$ and $R_{\mu}^{(1)}$ get shifted. This leads to  
\begin{align}
S_2^W&=-\frac{\kappa}{g_5^2z_0^2}\int d^4 x \langle u^{\mu}\ell_{\mu}^W\rangle+{\cal{O}}(\kappa^2)=\kappa \zeta_8\frac{f_{\pi}^4}{2}\int d^4 x \langle \Delta u_{\mu}u^{\mu}\rangle~,\label{O2}
\end{align}
which reproduces the chiral electroweak Lagrangian of Eq.~(\ref{chiral}) once one identifies\footnote{It is also common to work with the parameters $c_2$ and $g_8$, which are related through $c_2=f_{\pi}^4G_8=\kappa f_{\pi}^4 g_8$.}
\begin{align}
G_8=\frac{1}{2}\kappa\zeta_8~.
\end{align}
Notice that Eq.~(\ref{O2}) could have also been obtained in terms of currents from the first term in Eq.~(\ref{curr}):
\begin{equation}
S_2^W=\kappa \frac{\delta S_2}{\delta l_{\mu}}\delta l_{\mu}^W=\kappa \int d^4x \langle J_{L\mu}\delta l_{\mu}^W\rangle=\kappa \zeta_8\frac{f_{\pi}^4}{4}\int d^4x \left\langle u^{\dagger}u_{\mu}u\left\{\lambda_6,u^{\dagger}u^{\mu}u\right\}\right\rangle~,
\end{equation}
which can be reduced to Eq.~(\ref{O2}).

\subsection{${\cal{O}}(p^4)$ operators in the chiral expansion}

Using a common notation we will parameterize the full set of operators as 
\begin{equation}
{\mathcal{L}}_4^{\chi}=\sum_i L_i{\mathcal{O}}_i+G_8f_{\pi}^2\sum_j N_j{\mathcal{O}}_j^W~,
\end{equation}
where ${\cal{O}}_i$ and ${\mathcal{O}}_j^W$ are strong and electroweak operators, respectively. We have listed the relevant ones in the Appendix.

The holographic contributions to this order can be extracted from the zero-mode piece of the fields in the second line of Eq.~(\ref{5Dgfaction}),
\begin{align}
 S = -\frac{1}{4 g_5^2} \int d^4 x \int_{0}^{z_0} \frac{dz}{z} \eta^{\mu\,\rho}\,\eta^{\nu\,\sigma}\langle F_{(L)\mu\nu} F_{(L)\rho\sigma}+ F_{(R)\mu\nu}
F_{(R)\rho\sigma}\rangle~.
\end{align}
A potential contribution in the first line of Eq.~(\ref{5Dgfaction}) identically cancels because there is no mixing between the pion and axial resonances. This means that the low energy couplings obtained in this model do not receive contributions from resonance exchange, but rather should be regarded as geometric terms. Resonance exchange effects will only start at ${\cal{O}}(p^6)$. This seems to be a generic feature of using vector representations for spin-1 fields~\cite{Bando:1987br,Ecker:1989yg} and contrasts with the antisymmetric tensor representation of resonances, where vector exchange is known to start already at ${\cal{O}}(p^4)$~\cite{Ecker:1988te}.

In order to simplify the matching with the operators listed in the Appendix it will be convenient to work with the combinations $F_{\pm}^{\mu\nu}=F_L^{\mu\nu}\pm F_R^{\mu\nu}$, with the field strengths defined as $F_{L\mu\nu}=\partial_{\mu}{L}_{\nu}-\partial_{\nu}{L}_{\mu}-i[{L}_{\mu},{L}_{\nu}]$. On the UV boundary we will adopt the notation $f_{\pm}^{\mu\nu}=f_L^{\mu\nu}\pm f_R^{\mu\nu}$, where the field strengths are defined accordingly as $f_{L\mu\nu}=\partial_{\mu}L_{\nu}^{(0)}-\partial_{\nu}L_{\mu}^{(0)}-i[L_{\mu}^{(0)},L_{\nu}^{(0)}]$. Similar expressions hold for the right-handed fields. It will also prove convenient to write the right and left-handed fields as 
\begin{align}
L_{\mu}(x,z)&=\alpha_{+}\bigg[L_{\mu}^{(0)}+\kappa \ell_{\mu}^W\bigg]+\alpha_{-}R_{\mu}^{(0)}=L_{\mu}^{(0)}+\kappa\alpha_+\ell_{\mu}^W+\alpha_-u_{\mu}~,\nonumber\\
R_{\mu}(x,z)&=\alpha_{-}\bigg[L_{\mu}^{(0)}+\kappa \ell_{\mu}^W\bigg]+\alpha_{+}R_{\mu}^{(0)}=R_{\mu}^{(0)}+\kappa\alpha_-\ell_{\mu}^W-\alpha_-u_{\mu}~,\label{convenient}
\end{align}
where we have defined
\begin{equation}
\alpha_{\pm}=\frac{1\pm\alpha(z)}{2}~,\qquad\qquad \alpha(z)=1-\frac{z^2}{z_0^2}~.
\end{equation}
It is then rather straightforward to obtain
\begin{align}
F_{+\mu\nu}&=f_{+\mu\nu}+\frac{i}{2}(1-\alpha^2)[u_{\mu},u_{\nu}]+\kappa\Big[\omega_{\mu\nu}^W+i\frac{\alpha^2}{2}([\ell_{\mu}^W,u_{\nu}]-[\ell_{\nu}^W,u_{\mu}])\Big]~,\\
F_{-\mu\nu}&=\alpha f_{-\mu\nu}+\kappa\alpha\Big[ \omega_{\mu\nu}^W+\frac{i}{2}([\ell_{\mu}^W,u_{\nu}]-[\ell_{\nu}^W,u_{\mu}])\Big]~,
\end{align}
where we have defined $\omega_{\mu\nu}^W=\nabla_{\mu}\ell_{\nu}^W-\nabla_{\nu}\ell_{\mu}^W$ and $\nabla_{\mu}$ is the chiral covariant derivative, $\nabla_{\mu}\,\cdot=\partial_{\mu}\cdot+[\Gamma_{\mu},\cdot]$. Eq.~(\ref{5Dgfaction}) is therefore proportional to the combination
\begin{align}
F_{+\mu\nu}F_{+}^{\mu\nu}+F_{-\mu\nu}F_{-}^{\mu\nu}&=f_{+\mu\nu}f_{+}^{\mu\nu}+\alpha^2 f_{-\mu\nu}f_{-}^{\mu\nu}-\frac{1}{4}(1-\alpha^2)^2[u_{\mu},u_{\nu}][u^{\mu},u^{\nu}]+i(1-\alpha^2)f_{+\mu\nu}[u^{\mu},u^{\nu}]\nonumber\\
&+\kappa\Bigg\{2(f_{+}^{\mu\nu}+\alpha^2f_{-}^{\mu\nu})\omega_{\mu\nu}^W+2i\alpha^2(f_{+}^{\mu\nu}+f_{-}^{\mu\nu})[\ell_{\mu}^W,u_{\nu}]\nonumber\\
&+i(1-\alpha^2)\omega_{\mu\nu}^W[u^{\mu},u^{\nu}]+\alpha^2(\alpha^2-1)[\ell_{\mu}^W,u_{\nu}][u^{\mu},u^{\nu}]\Bigg\}~,\label{p4}
\end{align}
where the first line corresponds to the ${\cal{O}}(p^4)$ strong chiral Lagrangian, and the term in brackets contains the electroweak contribution. 

Comparing the first line of Eq.~(\ref{p4}) with the strong operators listed in the Appendix one finds the following predictions for the low energy coefficients of the strong sector~\cite{Hirn:2005nr}: 
\begin{align}
L_1 &= \displaystyle\frac{1}{32 g_5^2} \int_0^{z_{0}} \frac{dz}{z} \left[1-{\alpha(z)}^2\right]^2~,\label{GL1}\\
L_{10}& = \displaystyle-\frac{1}{4 g_5^2} \int_0^{z_0} \frac{dz}{z} \left[1-{\alpha(z)}^2\right]~,\label{GL10}\\
H_1 & = \displaystyle-\frac{1}{8 g_5^2} \int_0^{z_{0}} \frac{dz}{z}
\left[1+{\alpha(z)}^2\right]~,\label{GLH1}\\
L_2 &= 2 L_1, \quad L_3 = -6 L_1, \quad L_9 =
-L_{10}~.\label{GL239}
\end{align}
The relations connecting $L_1, L_2$ and $L_3$ are a direct consequence of the Skyrme structure of the only pure pion term, {\it{i.e.}}, $\langle[u_{\mu},u_{\nu}][u^{\mu},u^{\nu}]\rangle$, combined with the Cayley-Hamilton relation for $n_f=3$: 
\begin{equation}
4\langle u_{\mu}u^{\mu}u_{\nu}u^{\nu}\rangle+2\langle u_{\mu}u_{\nu}u^{\mu}u^{\nu}\rangle-2\langle u_{\mu}u^{\mu}\rangle^2-\langle u_{\mu}u_{\nu}\rangle^2=0~.
\end{equation}

Regarding the electroweak sector, the specific form for $\ell_{\mu}^W$ yields\footnote{Terms proportional to the identity matrix in $\omega_{\mu\nu}^W$ give vanishing contributions to Eq.~(\ref{p4}) and are therefore omitted.} 
\begin{equation}
\omega_{\mu\nu}^W=\zeta_8\frac{f_{\pi}^2}{2}\Bigg[\left\{\Delta,f_{-\mu\nu}\right\}+\left\{\nabla_{\nu}\Delta,u_{\mu}\right\}-\left\{\nabla_{\mu}\Delta,u_{\nu}\right\}\Bigg]~.
\end{equation}
Using this result and after a lengthy but straightforward derivation, the corresponding matching of Eq.~(\ref{p4}) to the electroweak operators yields the following results for the weak low-energy couplings:
\begin{align}\label{weakN}
N_1&=-\displaystyle\frac{1}{4 g_5^2} \int_0^{z_{0}} \frac{dz}{z}\left[(1-\alpha^2)\left(1-\frac{5}{3}\alpha^2\right)\right]~, &N_{16}&=2N_{18}=-\displaystyle\frac{1}{4 g_5^2} \int_0^{z_{0}} \frac{dz}{z}(1-3\alpha^2)~,\nonumber\\
N_2&=\displaystyle\frac{1}{4 g_5^2} \int_0^{z_{0}} \frac{dz}{z}\left[(1-\alpha^2)\left(1+\frac{7}{3}\alpha^2\right)\right]~, &N_{17}&=N_{20}=0~,\nonumber\\
N_3&=\frac{3}{2}N_4=-\displaystyle\frac{1}{2 g_5^2} \int_0^{z_{0}} \frac{dz}{z}\left[(1-\alpha^2)\alpha^2\right]~, &N_{25}&=\displaystyle\frac{1}{2 g_5^2} \int_0^{z_{0}} \frac{dz}{z}~,\nonumber\\
N_{14}&=-2N_{37}=\displaystyle\frac{1}{4 g_5^2} \int_0^{z_{0}} \frac{dz}{z}(1+\alpha^2)~, &N_{26}&=\displaystyle\frac{1}{2 g_5^2} \int_0^{z_{0}} \frac{dz}{z}\alpha^2~,\nonumber\\ 
N_{15}&=2N_{19}=\displaystyle\frac{1}{2 g_5^2} \int_0^{z_{0}} \frac{dz}{z}(1-\alpha^2)~, &N_{27}&=-\displaystyle\frac{1}{8g_5^2} \int_0^{z_{0}} \frac{dz}{z}(\alpha^2-1)~.
\end{align}
Similar to what happened in the strong sector, the matching of operators requires to use a Cayley-Hamilton relation for $n_f=3$, namely
\begin{equation}
\langle \Delta u_{\mu}u^{\mu}u_{\nu}u^{\nu}\rangle+3\langle \Delta u_{\mu}u_{\nu}u^{\mu}u^{\nu}\rangle+2\langle \Delta u_{\mu}u_{\nu}u^{\nu}u^{\mu}\rangle-3\langle\Delta u_{\mu}u_{\nu}\rangle\langle u^{\mu}u^{\nu}\rangle-2\langle \Delta u_{\mu}\rangle\langle u^{\mu}u_{\nu}u^{\nu}\rangle=0~.
\end{equation}


\subsection{Connection with WDM and factorization}
In principle, the analysis of the ${\cal{O}}(p^4)$ terms could have been performed in terms of the chiral currents by considering Eq.~(\ref{curr}), in a way analogous to what we did for the ${\cal{O}}(p^2)$. Thus, the strong and weak chiral operators could have also been obtained from
\begin{equation}\label{abst}
S_4=\frac{\delta S_4}{\delta r_{\mu}}r_{\mu}^{(0)}+\frac{\delta S_4}{\delta l_{\mu}}\left[l_{\mu}^{(0)}+\kappa \delta l_{\mu}^W\right]=\int d^4x \left[\langle J_{L\mu}^{(3)} l^{\mu\,(0)}+J_{R\mu}^{(3)} r^{\mu\,(0)}\rangle+\kappa \langle J_{L\mu}^{(3)}\delta l^{\mu W}\rangle\right]~.
\end{equation}
In practice, however, the determination of $J_{L\mu}^{(3)}$ and $J_{R\mu}^{(3)}$ turns out to be involved, and it is preferable to shift the strong Lagrangian as we did in the previous Section. Eq.~(\ref{abst}) is however useful from a formal standpoint. For instance, one readily sees that, as a consequence of the induced shift in the left-handed sources, there will be a direct relation between the strong and weak low-energy couplings. 

Different strategies have been previously considered in the literature to relate the weak couplings to the strong ones. The reason is that while chiral symmetry predicts the form of the strong and weak operators, experimental information is too scarce to constrain all the low-energy couplings. In order to be predictive, an extra mechanism has to be invoked. The most prominent approaches considered so far were the Factorization Model (FM)~\cite{Pich:1990mw} and the Weak Deformation Model (WDM)~\cite{EckerWDM}. The FM assumes that the weak current-current operators are, to a very good approximation, a product of color-singlet currents, {\it{i.e.}},  
\begin{align}\label{fact}
{\cal{L}}_{FM}&\sim \int d^4x \left\langle \lambda_6 \frac{\delta S}{\delta l_{\mu}}\frac{\delta S}{\delta l^{\mu}}\right\rangle=\int d^4x \left\langle \lambda_6 \left[\frac{\delta S_2}{\delta l_{\mu}}+\frac{\delta S_4}{\delta l_{\mu}}+\cdots\right]\left[\frac{\delta S_2}{\delta l^{\mu}}+\frac{\delta S_4}{\delta l^{\mu}}+\cdots\right]\right\rangle\nonumber\\
&=\int d^4x \left\langle \lambda_6 \left\{\frac{\delta S_2}{\delta l_{\mu}}\frac{\delta S_2}{\delta l^{\mu}}+\left[\frac{\delta S_2}{\delta l_{\mu}}\frac{\delta S_4}{\delta l^{\mu}}+\frac{\delta S_4}{\delta l_{\mu}}\frac{\delta S_2}{\delta l^{\mu}}\right]+\left[\frac{\delta S_2}{\delta l_{\mu}}\frac{\delta S_6}{\delta l^{\mu}}+\frac{\delta S_4}{\delta l_{\mu}}\frac{\delta S_4}{\delta l^{\mu}}+\frac{\delta S_6}{\delta l_{\mu}}\frac{\delta S_2}{\delta l^{\mu}}\right]+\cdots\right\}\right\rangle.
\end{align}
Corrections to the previous expression appear as gluon exchanges between the currents. The FM therefore assumes that they are a subleading effect.  

On the other hand, the WDM is based on the heuristic observation that the $O(p^2)$ weak Lagrangian can be obtained from the strong one by the substitution rule
\begin{equation}
u_\mu\rightarrow u_\mu+G_8f_{\pi}^2\left[\{u_\mu, \Delta\}-\frac{2}{3}\epsilon\langle u_\mu \Delta\rangle \textbf{1}_3\right]~.\label{WDMumu}
\end{equation}
The corresponding transformation of the chiral connection is achieved by requiring that the deformation is purely left-handed. This entails that
\begin{equation}
\Gamma_\mu\rightarrow \Gamma_\mu+\frac{i}{2}G_8f_{\pi}^2\left[\{u_\mu,\Delta\}-\frac{2}{3}\epsilon\langle u_\mu\Delta\rangle\textbf{1}_3\right]~,\label{WDMGamma}
\end{equation}
such that the right-handed combination $\Gamma_\mu-\frac{i}{2}u_{\mu}$ is left unchanged.

It is instructive at this point to examine the relation between the holographic model (HEW), WDM and FM. On the one hand, one can easily check that the induced shift of left-handed sources in HEW can be written in terms of vector and axial sources as
\begin{align}
V_{\mu}^{(0)}&=2i\Gamma_{\mu}+\kappa \ell_{\mu}^W~,\nonumber\\
A_{\mu}^{(0)}&=u_{\mu}-\kappa \ell_{\mu}^W~.
\end{align}
Using Eq.~(\ref{definition}) for $\ell_{\mu}^W$, together with $G_8=\kappa \zeta_8/2$, one concludes that the holographic deformation is equivalent to the WDM for $\epsilon=\frac{3}{2}$. On the other hand, the double-trace formalism we have employed assumes that the weak currents are color singlets, and therefore there has to be also a connection with FM. Actually, one can show that 
\begin{align}\label{HEW}
{\cal{L}}_{HEW}&\sim \frac{1}{2}\int d^4x \left\langle \frac{\delta S}{\delta l_{\mu}}\ell_{\mu}^W\right \rangle=\frac{1}{2}\int d^4x \left \langle\left(\frac{\delta S_2}{\delta l_{\mu}}+\frac{\delta S_4}{\delta l^{\mu}}+\frac{\delta S_6}{\delta l_{\mu}}+\cdots\right)\ell_{\mu}^W\right\rangle\nonumber\\
&=\int d^4x \left\langle \lambda_6 \left\{\frac{\delta S_2}{\delta l_{\mu}}\frac{\delta S_2}{\delta l^{\mu}}+\frac{1}{2}\left[\frac{\delta S_2}{\delta l_{\mu}}\frac{\delta S_4}{\delta l^{\mu}}+\frac{\delta S_4}{\delta l_{\mu}}\frac{\delta S_2}{\delta l^{\mu}}\right]+\frac{1}{2}\left[\frac{\delta S_2}{\delta l_{\mu}}\frac{\delta S_6}{\delta l^{\mu}}+\frac{\delta S_6}{\delta l_{\mu}}\frac{\delta S_2}{\delta l^{\mu}}\right]+\cdots\right\}\right\rangle,
\end{align}
where in the last line we have used that $\ell_{\mu}^W=\left\{\lambda_6,\frac{\delta S_2}{\delta l_{\mu}}\right\}$. Direct comparison with Eq.~(\ref{fact}) shows that FM and HEW are equivalent to leading order, while at subleading order they differ only by a numerical factor. Beyond that order HEW has missing terms.

It is instructive to compare our previous result with the statement made in Ref.~\cite{Ecker:1992de}, namely that WDM and FM are equivalent up to ${\cal{O}}(p^4)$, if one corrects for a fudge factor $\frac{1}{2}$. We observe that Eq.~(\ref{HEW}) indeed complies with this statement. However, it would be more accurate to state that WDM is a truncated version of factorization. The same applies to HEW. This can be easily seen if we let $\ell_{\mu}^W=\left\{\lambda_6,\frac{\delta S_2}{\delta l_{\mu}}+\frac{\delta S_4}{\delta l_{\mu}}\right\}$. Then one can readily see that the fudge factor between both models is no longer needed. Only now one can claim that WDM and factorization are equivalent up to ${\cal{O}}(p^4)$. By induction one can conclude that both models will be equivalent only when the sources contain the resummed chiral current, {\it{i.e.}}, $\ell_{\mu}^W=\left\{\lambda_6,\frac{\delta S}{\delta l_{\mu}}\right\}$. In the conclusions we will comment on the possibility of enhancing the chiral current this way in HEW. 

Strictly speaking, the previous connections between models are only valid if the $\epsilon$-term in Eqs.~(\ref{WDMumu}) and (\ref{WDMGamma}) is ignored. We already discussed that WDM and HEW are equivalent for the particular value $\epsilon=\frac{3}{2}$. The $\epsilon$-term is proportional to the identity and therefore related to the presence (or absence) of singlet sources. In the absence of singlet sources, consistency in the WDM requires that $\epsilon=1$, but $\epsilon$ is left undetermined once singlet sources are allowed in. Therefore, HEW is equivalent to WDM up to terms involving singlet sources. On the other hand, in Ref.~\cite{Ecker:1992de} it was shown that the above-mentioned equivalence between WDM and FM at ${\cal{O}}(p^4)$ persisted in the presence of singlet sources only if $\epsilon=\frac{3}{2}$. 

At ${\cal{O}}(p^4)$ one can show that the $\epsilon$-term is only relevant in the parity-odd sector (see next Section). Therefore, the weak couplings we found in Eq.~(\ref{weakN}) are $\epsilon$-blind and as a result they can be shown to satisfy the WDM relations reported in~\cite{Ecker:1992de}:
\begin{align}
N_1&=-\frac{40}{3}L_1+\frac{2}{3}L_9~,&\qquad N_4&=\displaystyle\frac{32}{3}L_1-\frac{4}{3}L_9~,&\qquad N_{16}&=-2L_9-2H_1~,\nonumber\\
N_2&=-\frac{56}{3}L_1+\frac{10}{3}L_9~,&\qquad N_{14}&=-2H_1~,&\qquad N_{17}&=0~,\\
N_3&=16L_1-2L_9~,& \qquad N_{15}&=2L_9~,&\qquad N_{18}&=-L_9-H_1~,\nonumber
\end{align}
which we complement with
\begin{align}
N_{19}&=L_9~,&\qquad N_{26}&=-L_9-2H_1~,\nonumber\\
N_{20}&=0~,&\qquad N_{27}&=\frac{1}{2}L_{9}~,\\
N_{25}&=L_9-2H_1~,&\qquad N_{37}&=H_1~.\nonumber
\end{align}
Since we are working in the strict large-$N_c$ limit, the previous relations are blind to the scale dependence of the chiral couplings. However, we want to remark that for those weak couplings only sensitive to vector and axial-vector exchange, {\emph{i.e.}}, $N_{14-18}, N_{25}, N_{27}$ and $N_{37}$, there is exact matching of the anomalous dimensions.\footnote{The renormalization of the chiral couplings can be found in Ref.~\cite{Gasser:1984gg} (strong sector) and in Ref.~\cite{Ecker:1992de} (weak sector).} The matching breaks down whenever a scalar contribution is expected. This breakdown is not surprising because we are not including scalars in our model. Thus, we conclude that the relations above involving only vector exchange are stable under renormalization and should be regarded accordingly as predictions valid to all orders in the chiral expansion.   


\section{The anomalous sector}\label{secV}
As pointed out in~\cite{Hill:2006wu}, anomalous 4-dimensional processes can be fully reproduced from the 5-dimensional Chern-Simons term. The procedure outlined in the previous sections can be straightforwardly generalized to the odd-parity sector of the theory. From a 5-dimensional point of view, one has to add the Chern-Simons form to the 5-dimensional action. Its expression for chiral theories is  
\begin{equation}
S_{\rm{CS}}=\frac{N_c}{24\pi^2}\int_{\mathrm{AdS_5}}\left[\omega_5(L_{M})-\omega_5(R_{M})\right]~,
\end{equation}
where
\begin{equation}
\omega_5(L)=\left\langle LF_{(L)}^2-\frac{1}{2}L^3F_{(L)}+\frac{1}{10}L^5\right\rangle~,\qquad F_{(L)}=dL+L^2~.
\end{equation}
Recall that the multiple-trace deformation only affects the source terms, {\it{i.e.}}, it is localized on the UV brane. In general the left and right-handed fields contain a source and a resonance term. However, in order to obtain the ${\cal{O}}(p^4)$ weak chiral operators one only needs to consider the source terms. With source terms the weak deformation and the integration over the fifth dimension commute. Thus, in order to obtain the weak odd-parity operators we just need to shift the left-handed sources in the resulting four-dimensional action, which is nothing but the gauged Wess-Zumino-Witten (WZW) action. In the following we will give some details of the calculation.

In Section~\ref{secII} we used the gauge freedom of the $L_M$ and $R_M$ fields to eliminate their fifth components through a chiral rotation, namely
\begin{align} 
L_M^{\xi}(x,z)&=\xi_L^\dagger(x,z)[L_M(x,z)+i\partial_M]\xi_L(x,z)~,\nonumber\\ R_M^{\xi}(x,z)&=\xi_R^\dagger(x,z)[R_M(x,z)+i\partial_M]\xi_R(x,z)~,
\end{align}
$\xi_{L,R}(x,z)$ being the Wilson lines defined in Eq.~(\ref{Wilson}). Contrary to the Yang-Mills term, the Chern-Simons term is not gauge invariant and it changes instead to
\begin{eqnarray}
S_{\rm{CS}}&=&\frac{N_c}{24\pi^2}\left\{\int_{\mathrm{M_5}}\left[\omega_5(L^{\xi})-\frac{1}{10}\langle (d\xi_L\xi_L^{\dagger})^5\rangle\right]-\int_{\partial M_5}\alpha_4(\sigma,L)\right\}-(L\leftrightarrow R)\label{CStrans1}~,
\end{eqnarray}
where $\sigma=du^{\dagger}u$, $(L\leftrightarrow R)$ stands for the replacements $(L;L^{\xi};\,\sigma)\to(R;R^{\xi};\,-u \sigma u^{\dagger})$ and $\alpha_4$ is the Bardeen counterterm
\begin{equation}
\alpha_4(\sigma,B)=-\frac{1}{2}\left\langle \sigma (BF_B+F_BB)-\sigma B^3-\frac{1}{2}\sigma B\sigma B-\sigma^3B\right\rangle~.
\end{equation}
While the second term in~(\ref{CStrans1}) is a genuine 5-dimensional object, $\omega_5$ can be integrated over the fifth dimension. Combining the integrated $\omega_5$ with $\alpha_4$, the Chern-Simons term can be shown to lead to the gauged  WZW action:
\begin{equation}
S_{CS}=-\frac{iN_c}{48\pi^2}\Bigg[\int d^4x~ \varepsilon_{\mu\nu\lambda\rho}\langle W[U,l,r]-W[1,l,r]\rangle^{\mu\nu\lambda\rho}-\frac{i}{5}\int_{\mathrm{AdS}_5} \langle \Sigma^5\rangle\Bigg]~,
\end{equation} 
where $\Sigma=U^{\dagger}dU=i\Sigma_{M}dx^{M}$ and
\begin{align}
W&[U,l,r]_{\mu\nu\lambda\rho}=l_{\mu}l_{\nu}l_{\lambda}\hat{r}_{\rho}+\frac{1}{4}l_{\mu}\hat{r}_{\nu}l_{\lambda}\hat{r}_{\rho}-i\Sigma_{\mu} l_{\nu}\hat{r}_{\lambda}l_{\rho}-i\Sigma_{\mu} l_{\nu}l_{\lambda}l_{\rho}+\frac{1}{2}\Sigma_{\mu} l_{\nu}\Sigma_{\lambda}l_{\rho}-\Sigma_{\mu}\Sigma_{\nu}\hat{r}_{\lambda}l_{\rho}\nonumber\\
&-i\Sigma_{\mu}\Sigma_{\nu}\Sigma_{\lambda}l_{\rho}+i\partial_{\mu}l_{\nu}l_{\lambda}\hat{r}_{\rho}
+i({\widehat{\partial_{\mu} r_{\nu}}})l_{\lambda}\hat{r}_{\rho}+\Sigma_{\mu}(\widehat{\partial_{\nu} r_{\lambda}})l_{\rho}+\Sigma_{\mu} l_{\nu}\partial_{\lambda} l_{\rho}+\Sigma_{\mu} \partial_{\nu} l_{\lambda} l_{\rho}-(l\leftrightarrow r)~,\label{WZW}
\end{align}
where now $(l\leftrightarrow r)$ stands for $(l_{\mu};\,\Sigma_{\mu})\to(r_{\mu};\,-U^{\dagger} \Sigma_{\mu} U)$. In Eq.~(\ref{WZW}) above we have absorbed the chiral field $U(x)$ using the short-hand notation $\hat{X}_{\mu} \to U^{\dagger}X_{\mu} U$ for all right-handed fields. If we now apply the usual shift in the left-handed sources this will result in $W[U,l,r]\to W[U,l,r]+\kappa\Delta[U,l,r]$, where the generated $\Delta[U,l,r]$ piece will contain the weak odd-parity operators. Notice that the shift also affects the term $W[1,l,r]$, and accordingly a term like $\Delta[1,l,r]$ is also generated. However, this piece only contains source terms and therefore will only contribute as contact terms. The expression for $\Delta[U,l,r]$ is given by
\begin{align}
\Delta&[U,l,r]_{\mu\nu\lambda\rho}=\ell_{\mu}^Wl_{\nu}l_{\lambda}\hat{r}_{\rho}+l_{\mu}\ell_{\nu}^W l_{\lambda}\hat{r}_{\rho}+l_{\mu}l_{\nu}\ell_{\lambda}^W\hat{r}_{\rho}+\ell_{\mu}^W\hat{r}_{\nu}l_{\lambda}\hat{r}_{\rho}-i\Sigma_{\mu}\ell_{\nu}^W\hat{r}_{\lambda}l_{\rho}-i\Sigma_{\mu}l_{\nu}\hat{r}_{\lambda}\ell_{\rho}^W\nonumber\\
&-i\Sigma_{\mu} \ell_{\nu}^W l_{\lambda}l_{\rho}-i\Sigma_{\mu}l_{\nu}\ell_{\lambda}^W l_{\rho}-i\Sigma_{\mu}l_{\nu}l_{\lambda}\ell_{\rho}^W+\Sigma_{\mu}\ell_{\nu}^W\Sigma_{\lambda}l_{\rho}-\Sigma_{\mu}\Sigma_{\nu}\hat{r}_{\lambda}\ell_{\rho}^W-i\Sigma_{\mu}\Sigma_{\nu}\Sigma_{\lambda}\ell_{\rho}^W+i\partial_{\mu}\ell_{\nu}^W l_{\lambda}\hat{r}_{\rho}\nonumber\\
&+i\partial_{\mu} l_{\nu}\ell_{\lambda}^W\hat{r}_{\rho}+i({\widehat{\partial_{\mu} r_{\nu}}})\ell_{\lambda}^W\hat{r}_{\rho}+\Sigma_{\mu}(\widehat{\partial_{\nu} r_{\lambda}})\ell_{\rho}^W+\Sigma_{\mu} \ell_{\nu}^W\partial_{\lambda}l_{\rho}+\Sigma_{\mu}l_{\nu}\partial_{\lambda}\ell_{\rho}^W+\Sigma_{\mu} \partial_{\nu} \ell_{\lambda}^W l_{\rho}+\Sigma_{\mu} \partial_{\nu}l_{\lambda}\ell_{\rho}^W\nonumber\\
&-\hat{r}_{\mu}\hat{r}_{\nu}\hat{r}_{\lambda}\ell_{\rho}^W-i\Sigma_{\mu}\hat{r}_{\nu}\ell_{\lambda}^W\hat{r}_{\rho}+\Sigma_{\mu}\Sigma_{\nu}\ell_{\lambda}^W\hat{r}_{\rho}-i(\widehat{\partial_{\mu}r_{\nu}})\hat{r}_{\lambda}\ell_{\rho}^W-i\partial_{\mu}\ell_{\nu}^W\hat{r}_{\lambda}l_{\rho}-i\partial_{\mu}l_{\nu}\hat{r}_{\lambda}\ell_{\rho}^W+\Sigma_{\mu}\partial_{\nu}\ell_{\lambda}^W\hat{r}_{\rho}~.
\end{align}
Matching the previous expression to the weak odd-parity operators listed in the Appendix can be eventually achieved using the explicit expression for $\ell_{\mu}^W$ and the useful relation $u_{\mu}={\hat{r}}_{\mu}-l_{\mu}+i\Sigma_{\mu}$. Alternatively, one could use Eq.~(\ref{curr}) in terms of the anomalous current. This turns out to be the shortest path to obtain the odd-parity operators. The reason is that, contrary to the even-parity operators, the topological structure of the WZW term allows to compute the associated current in a very compact way as the 3-form:\footnote{We are omitting the contact terms coming from $W[1,l,r]$ which, as we argued before, are irrelevant for our discussion.}
\begin{align}
J^L_{an}&=S_{WZW}\frac{\overleftarrow{\delta}}{\delta l}=-\frac{N_c}{48\pi^2}\left(i({\cal{L}})^3+\left\{{\cal{L}},F_{(L)}+\frac{1}{2}{\hat{F}}_{(R)}\right\}\right)~,
\end{align}
where ${\cal{L}}=iU^{\dagger}DU$.  The weak chiral operators follow from $S^{(W)}_{4}=\langle J^L_{an}{\delta l^W}\rangle$, and it is a rather straightforward exercise to conclude that
\begin{align}\label{anomalous}
N_{28}=\frac{2}{3}N_{30}=2N_{31}&=\frac{N_C}{48\pi^2}~,\nonumber\\
N_{29}=\frac{N_{32}}{2}=\frac{N_{33}}{2}=N_{34}&=\frac{N_C}{192\pi^2}~.
\end{align}
Due to the topological nature of the Chern-Simons term the previous expressions do not depend on the details of an underlying theory of hadrons. Most of the relations were first found in~\cite{Cheng:1990wx} in the study of radiative kaon decays in the factorization limit and later derived from more general arguments~\cite{Bijnens:1992ky}. However, the overall normalization of the $N_i$ found above within HEW differs from the factorization predictions by the usual $1/2$ factor we discussed in the previous Section. Additionally, we want to note that this agreement is highly nontrivial: the relations in the first line above depend crucially on the choice $\epsilon=\frac{3}{2}$, which, as we showed in the previous Section, is a genuine prediction of {\emph{both}} HEW and FM.     

\section{The role of vector mesons in $K\to 3\pi$ decays}\label{secVI}

We have already discussed that one of the underlying assumptions in our holographic treatment of the weak interactions is that vector resonance exchange is the dominant contribution to both strong and weak chiral couplings. This assumption was shown to work extremely well for the strong sector~\cite{Ecker:1988te} and for many weak-interacting processes. A singular exception was $K\to 3\pi$ decays: different hadronic models based on VMD~\cite{Isidori:1991ya,Ecker:1992de,D'Ambrosio:1997tb} found that not only vector meson exchange was not dominant, but turned out to vanish. The fact that different models reached the same conclusion was taken as evidence that vector meson saturation was failing in those particular channels. This however posed a two-fold puzzle: first, what made $K\to 3\pi$ decays so exceptional remained unexplained; and second, without the vector contributions, theoretical predictions were hard to reconcile with experimental data. In this Section we will show that the cancellation found in VMD models was accidental and based on a predicted relation between strong low-energy couplings $L_i$ not supported by phenomenology. In contrast, the corresponding relation between $L_i$ in the holographic model {\emph{does}} comply with phenomenology. This alone yields a nonvanishing vector exchange contribution to $K\to 3\pi$ quite in good agreement with experiment.  

Let us parametrize the $K\to 3\pi$ amplitudes, following Ref.~\cite{Kambor:1991ah}, as\footnote{For consistency we are only retaining the dominant octet contributions.} 
\begin{eqnarray}\label{k3pi}
{\cal{M}}(K_L\to \pi^+\pi^-\pi^0)&=&\alpha_1-\beta_1u+(\zeta_1+\xi_1)u^2+\frac{1}{3}(\zeta_1-\xi_1)v^2~,\nonumber\\
{\cal{M}}(K_L\to \pi^0\pi^0\pi^0)&=&-3\alpha_1-\zeta_1(3u^2+v^2)~,\nonumber\\
{\cal{M}}(K^+\to \pi^+\pi^+\pi^-)&=&2\alpha_1+\beta_1u+(2\zeta_1-\xi_1)u^2+\frac{1}{3}(2\zeta_1+\xi_1)v^2~,\nonumber\\
{\cal{M}}(K^+\to \pi^+\pi^0\pi^0)&=&-\alpha_1+\beta_1u-(\zeta_1+\xi_1)u^2-\frac{1}{3}(\zeta_1-\xi_1)v^2~,
\end{eqnarray}
where
\begin{equation}
u=\frac{s_3-s_0}{m_{\pi}^2}~,\qquad v=\frac{s_1-s_2}{m_{\pi}^2}~,\qquad  s_i=(p_K-p_{\pi_i})^2~,\qquad s_0=\frac{1}{3}\sum_{i=1}^3 s_i~.
\end{equation}
The amplitudes have been computed up to ${\cal{O}}(p^4)$ in ChPT~\cite{Kambor:1991ah}, giving the results
\begin{eqnarray}
\alpha_1&=&\alpha_1^{(0)}-\frac{2g_8}{27f_Kf_{\pi}}m_K^4\left\{(k_1-k_2)+24{\cal{L}}_1\right\}~,\nonumber\\
\beta_1&=&\beta_1^{(0)}-\frac{g_8}{9f_Kf_{\pi}}m_{\pi}^2m_K^2\left\{(k_3-2k_1)-24{\cal{L}}_2\right\}~,\nonumber\\
\zeta_1&=&-\frac{g_8}{6f_Kf_{\pi}}m_{\pi}^4\left\{k_2-24{\cal{L}}_1\right\}~,\nonumber\\
\xi_1&=&-\frac{g_8}{6f_Kf_{\pi}}m_{\pi}^4\left\{k_3-24{\cal{L}}_2\right\}~,
\end{eqnarray} 
where ${\cal{L}}_1={\cal{L}}_2+3L_2=2L_1+2L_2+L_3$ come from diagrams containing strong amplitudes with weak external vertices, while $k_1=9(-N_5+2N_7-2N_8-N_9)$, $k_2=3(N_1+N_2+2N_3)$ and $k_3=3(N_1+N_2-N_3)$ collect the direct weak terms. The vector meson exchange contributions to $k_i$ within the factorization model were computed in~\cite{Isidori:1991ya} and later extended to the scalar sector~\cite{Ecker:1992de}. On general grounds, $k_1$ only receives contributions from scalar mesons, while $k_2$ and $k_3$ also include vectors. It was already pointed out in~\cite{EckerWDM} that strong cancellations between the strong and weak contributions are to be expected. However, one of the puzzles of the factorization model in $K\to 3\pi$ was the apparent failure of vector meson dominance: the vector contributions to $k_2$ and $k_3$ identically cancelled, in contradiction with fits to experimental data~\cite{Kambor:1991ah,Devlin:1978ye}. 

Here we will reexamine this issue from the holographic electroweak model. One can show that in both the WDM and the holographic approach the vector contributions satisfy 
\begin{align}
&k_2=24{\cal{L}}_1~,\nonumber\\
&k_3=24\left({\cal{L}}_2+\frac{3}{4}L_9\right)~.\label{predictions}
\end{align}
Using Eqs.~(\ref{GL239}) one can further show that ${\cal{L}}_1=0$ and ${\cal{L}}_2=L_3$. This results come from the Skyrme structure of the ${\cal{O}}(p^4)$ Lagrangian and again it is common to both the WDM and the holographic model. This means that $\alpha^V_1=\zeta_1=0$ while $\beta_1,\xi_1\sim (k_3-24L_3)$.

One can now compare the previous predictions with the latest experimental results~\cite{Batley:2000zz,:2008js}. On the one hand, $\zeta_1$ has been shown to be compatible with zero in the neutral channel $K_L\to \pi^0\pi^0\pi^0$~\cite{:2008js}, a result that is confirmed in the charged channel~\cite{Batley:2000zz}. However, final state interactions are sizeable enough to preclude solid conclusions: in the neutral channel there are large uncertainties associated with rescattering effects~\cite{D'Ambrosio:1994km}, while the charged channel is very sensitive to cusp effects~\cite{Cabibbo:2005ez}. 

On the other hand, $\beta_1$ and even $\xi_1$ seem to be distinctly different from zero. Fits to experimental data suggest that $k_3\sim 5\cdot 10^{-9}$, with an error difficult to determine but roughly estimated around $30\%$. In models with VMD~\cite{Isidori:1991ya,Ecker:1992de,D'Ambrosio:1997tb}, $L_3=-\frac{3}{4}L_9$ and therefore Eq.~(\ref{predictions}) yields $k_3=0$, implying that vector contributions come entirely from strong vertices. As mentioned above, this result not only contradicts the fits but introduces a conceptual hurdle. However, in the holographic model $L_3=-\frac{11}{24}L_9$ and hence $k_3\sim 3\cdot 10^{-9}$. This not only stands in better agreement with experiment (it corresponds to a $50\%$ enhancement on both $\beta_1$ and $\xi_1$) but, since $k_3\neq 0$, it shows that the cancellation found in the literature was not generic but a model-dependent artefact of VMD models. Actually, we want to emphasize that the holographic prediction between $L_3$ and $L_9$ turns out to be much closer to the accepted phenomenological values~\cite{Ecker:1988te} than the VMD one. In other words, a better determination of the strong couplings in Eq.~(\ref{predictions}) naturally brings a nonvanishing vector meson contribution to $K\to 3\pi$ decays, such that predictions get closer to the experimental values.   
  

\section{Conclusions}\label{secVII}

In this paper we have applied the formalism of multiple-trace operators in holography to incorporate weak-interacting phenomena to a Yang-Mills--Chern-Simons holographic action. In this model, the pion field is realized in a non-linear way, thereby allowing a direct connection with chiral perturbation theory. This connection is worked out in detail by deriving the chiral Lagrangian for the strong and electroweak sectors up to next-to-leading order in the chiral expansion. Definite predictions for the low-energy couplings are given, both for the even and odd-parity sectors. The former follow from the Yang-Mills term while the latter stem from the Chern-Simons action.

One of the interesting consequences of the double-trace formalism is that the strong and weak couplings are related. In physical terms, this means that the underlying physical assumption of the model is that the effects that saturate the strong couplings are also responsible for the weak couplings. Since holographic models are models of hadronic resonances, we are implicitly assuming that all low energy couplings are determined by resonance saturation, {\emph{i.e.}}, that pure weak short distance effects are negligible. However, resonance effects only influence low-energy couplings starting at ${\cal{O}}(p^6)$. At ${\cal{O}}(p^4)$ the chiral couplings arise purely as geometric terms, {\it{i.e.}}, five-dimensional integrals in terms of the AdS metric alone.

The results of the model are then used to reexamine the $K\to 3\pi$ decays. Our conclusion is that the claim made in the past by different groups that vector meson dominance fails there is not generic. Rather, the cancellations predicted by different hadronic models turn out to be based on a constraint between the strong chiral couplings $L_3$ and $L_9$ that is not supported by phenomenology. In contrast, the holographic model gives a non-zero vector meson contribution precisely because the constraints in the strong sector are in much better agreement with phenomenology. In summary, the puzzling vector exchange cancellation in $K\to 3\pi$ turns out to be a model-dependent fine-tuning artefact. Beyond the phenomenological impact that this might have (most of the parameters are poorly determined because of strong final state interactions) we believe that it settles a conceptual issue.

Another interesting aspect of the holographic prescription adopted in this work is that HEW can be viewed as the five-dimensional analog of the heuristic WDM model introduced in Ref.~\cite{EckerWDM} (up to singlet source terms). Schematically,
\begin{align}
{\cal{L}}_{HEW}&\sim \frac{1}{2}\int d^4x \left \langle\left(\frac{\delta S_2}{\delta l_{\mu}}+\frac{\delta S_4}{\delta l_{\mu}}+\frac{\delta S_6}{\delta l_{\mu}}+\cdots\right)\ell_{\mu}^W\right\rangle~,\qquad \frac{\delta S_i}{\delta l^{\mu}}=J_{L\mu}^{(i-1)}~,
\end{align}
where $J_{L\mu}^{(i-1)}$ are the strong currents in the chiral expansion. Notice that this equivalence between WDM and HEW is highly nontrivial: in the WDM, $J_{L\mu}^{(i-1)}$ is taken from the strong chiral Lagrangian, while $\ell_{\mu}^W$ is determined from a heuristic prescription. In the HEW, $J_{L\mu}^{(i-1)}$ comes from the bulk action, while $\ell_{\mu}^W$ comes from a double-trace perturbation on the UV boundary. 

We have shown that both models are a truncated version of factorization. In principle factorization could be recovered order by order if the left-handed shift is extended beyond the leading term as $\ell_{\mu}^W=\left\{\lambda_6,\frac{\delta S_2}{\delta l^{\mu}}+\frac{\delta S_4}{\delta l^{\mu}}+\cdots\right\}$. The way this is achieved differs for each model: in WDM one needs to provide an {\emph{ad hoc}} shift in the $u_{\mu}$ and $\Gamma_{\mu}$ fields for each order in the chiral expansion. On the holographic side, following the prescription for double-trace operators, both the currents and left-handed sources should acquire extra pieces
\begin{align}
\langle J_{L\,\mu}(x)\rangle_{l_\mu}&=\frac{\delta
W_0[J_{L\mu},J_{R\mu}]}{\delta l^\mu(x)}=J_{L\,\mu}^{(1)}(x)+J_{L\,\mu}^{(3)}(x)+\cdots~,\label{hipo1}\\
l_\mu&\rightarrow l_{\mu}+\kappa \frac{\delta W_1[J_{L\mu}]}{\delta J_{L}^{\mu}}=l_\mu +\kappa \ell_{\mu\,(1)}^W+\kappa \ell_{\mu\,(3)}^W+\cdots~,\label{hipo2}
\end{align}
However, since the boundary action is entirely given by the (universal) ${\cal{O}}(p^2)$ pion terms, the left-handed sources are shifted only with the associated $J_{L\mu}^{(1)}$ chiral current. This seems to be a characteristic of the double-trace formalism. 

To illustrate this point, consider the generalized action 
\begin{align}
S_5&=\int d^4x\int_0^{z_0} dz \sqrt{g}\left\langle\alpha_2 F^2+\alpha_3 F^3+\alpha_4 F^4+\cdots\right\rangle~,
\end{align}
where the first piece is the Yang-Mills term, and the higher-order operators allow to go beyond ${\cal{O}}(p^4)$. Notice however that those extra bulk operators cannot contribute to Eq.~(\ref{hipo1}). To see this consider the boundary action, which will be modified to
\begin{align}
W_0[J_{L\mu},J_{R\mu}]\sim\int d^4x\left\langle L^{(0)}_{\mu}(x)L^{(1)}_{\nu}(x)\left[\eta^{\mu\nu}+\lambda_3(z) F^{\mu\nu}+\cdots\right]\right\rangle\bigg|_{z\to 0}~.
\end{align} 
It can be shown that the coefficient $\lambda_3(z)\sim \alpha_3 z^2$ and therefore vanishes. Positive powers of $z$ will generically appear as metric factors associated with higher-order operators. Thus, no NLO contributions can be generated in Eqs.~(\ref{hipo1}) and (\ref{hipo2}). We suspect that this conclusion goes beyond the double-trace prescription and is a generic built-in feature of holographic models of QCD.  


\section*{Acknowledgements}
O.~C.~wants to thank the University of Naples for very pleasant stays during the different stages of this work. L.~C.~ and G.~D'A.~ are supported in part by MIUR, Italy, under project 2005-023102 and by Fondo Dipartimentale per la Ricerca 2009. O.~C.~is supported by MICINN (Spain) under Grants FPA2007-60323, AIC10-D-000591 and by the Spanish Consolider Ingenio 2010 Programme CPAN (CSD2007-00042).\\
  

\section*{Appendix}
We will define the operators entering the ${\cal{O}}(p^4)$ chiral Lagrangian as
\begin{equation}
{\mathcal{L}}_4^{\chi}=\sum_i L_i{\mathcal{O}}_i+G_8f_{\pi}^2\sum_j N_j{\mathcal{O}}_j^W~,
\end{equation}
where the first term collects the strong sector and the second the weak sector, including both odd and even-parity operators. For the strong sector we will use the original basis of Ref.~\cite{Gasser:1984gg}: 
\begin{align}
{\cal{O}}_1&=\langle u_{\mu}u^{\mu}\rangle^2~,& {\cal{O}}_9&=-i\langle f_{+}^{\mu\nu}u_{\mu}u_{\nu}\rangle~,\nonumber\\
{\cal{O}}_2&=\langle u_{\mu}u^{\nu}\rangle^2~,& {\cal{O}}_{10}&=\frac{1}{4}\langle f_{+\mu\nu}f_+^{\mu\nu}-f_{-\mu\nu}f_-^{\mu\nu}\rangle~,\nonumber\\
{\cal{O}}_3&=\langle u_{\mu}u^{\mu}u_{\nu}u^{\nu}\rangle~,& {\cal{O}}_{11}&=\frac{1}{2}\langle f_{+\mu\nu}f_+^{\mu\nu}+f_{-\mu\nu}f_-^{\mu\nu}\rangle~,
\end{align}
where we identify $L_{11}\equiv H_1$. Notice that we are only including those operators which generate the low energy dynamics of vector and axial-vector modes, and disregarding scalar and pseudoscalar effects.  

Concerning the electroweak sector, we will adopt the basis employed in~\cite{Ecker:1992de}:
\begin{equation}
\begin{array}{ll}
{\mathcal{O}}_1^W=\langle \Delta u_{\mu}u^{\mu}u_{\nu}u^{\nu}\rangle~, & {\mathcal{O}}_{28}^W=i\epsilon_{\mu\nu\lambda\rho}\langle\Delta u^{\mu}\rangle\langle u^{\nu}u^{\lambda}u^{\rho}\rangle~,\\
{\mathcal{O}}_2^W=\langle \Delta u_{\mu}u_{\nu}u^{\nu}u^{\mu}\rangle~,&
{\mathcal{O}}_{29}^W= \epsilon_{\mu\nu\lambda\rho}\langle\Delta[f_{+}^{\mu\nu}-f_{-}^{\mu\nu},u^{\lambda}u^{\rho}]\rangle~,\\
{\mathcal{O}}_3^W=\langle \Delta u_{\mu}u_{\nu}\rangle\langle u^{\mu}u^{\nu}\rangle~,&
{\mathcal{O}}_{30}^W= \epsilon_{\mu\nu\lambda\rho}\langle\Delta u^{\mu}\rangle\langle f_+^{\lambda\rho} u^{\nu}\rangle~,\\
{\mathcal{O}}_4^W=\langle \Delta u_{\mu}\rangle\langle u^{\mu}u_{\nu}u^{\nu}\rangle~,&
{\mathcal{O}}_{31}^W= \epsilon_{\mu\nu\lambda\rho}\langle\Delta u^{\mu}\rangle\langle f_-^{\lambda\rho} u^{\nu}\rangle~,\\
{\mathcal{O}}_{14}^W=i\langle \Delta \left\{f_{+\mu\nu},u^{\mu}u^{\nu}\right\}\rangle~,&
{\mathcal{O}}_{32}^W= i\epsilon_{\mu\nu\lambda\rho}\langle\hat\nabla^{\mu}\Delta[f_{+}^{\lambda\rho},u^{\nu}]\rangle~,\\
{\mathcal{O}}_{15}^W=i\langle \Delta u^{\mu}f_{+\mu\nu}u^{\nu}\rangle~,&
{\mathcal{O}}_{33}^W= i\epsilon_{\mu\nu\lambda\rho}\langle\hat\nabla^{\mu}\Delta[f_{-}^{\lambda\rho},u^{\nu}]\rangle~,\\
{\mathcal{O}}_{16}^W=i\langle \Delta \left\{f_{-\mu\nu},u^{\mu}u^{\nu}\right\}\rangle~,&
{\mathcal{O}}_{34}^W= \epsilon_{\mu\nu\lambda\rho}\langle\Delta[f_{+}^{\mu\nu}+f_{-}^{\mu\nu},u^{\lambda}u^{\rho}]\rangle~,\\
{\mathcal{O}}_{17}^W=i\langle \Delta u^{\mu}f_{-\mu\nu}u^{\nu}\rangle~,&
{\mathcal{O}}_{35}^W= i\epsilon_{\mu\nu\lambda\rho}\langle\Delta[f_{+}^{\lambda\rho},f_{-}^{\mu\nu}]\rangle~,\\
{\mathcal{O}}_{18}^W=\langle \Delta (f_{+\mu\nu}f_{+}^{\mu\nu}-f_{-\mu\nu}f_{-}^{\mu\nu})\rangle~,&\\
{\mathcal{O}}_{27}^W=\langle \Delta (2 f_{+\mu\nu}f_{+}^{\mu\nu}-\left\{f_{+\mu\nu},f_{-}^{\mu\nu}\right\}\rangle~,&\\
{\mathcal{O}}_{37}^W=\langle \Delta (f_{+\mu\nu}+f_{+\mu\nu})(f_{+}^{\mu\nu}+f_{-}^{\mu\nu})\rangle~,&
\end{array}
\end{equation}
where the left-hand (right-hand) side collects the even-parity (odd-parity) operators and the definition $\hat\nabla_{\mu}\Delta=\nabla_{\mu}\Delta+\frac{i}{2}[u_{\mu},\Delta]$ has been used.

\end{document}